\documentclass[conference]{IEEEtran}
\IEEEoverridecommandlockouts

\ifCLASSINFOpdf
\else
\fi

\usepackage[T1]{fontenc}  
\usepackage{lmodern}      
\usepackage{times}        

\usepackage[a4paper, left=0.625in, right=0.625in, top=0.75in, bottom=1.0in]{geometry}

\usepackage{caption}
\usepackage{subcaption}
\usepackage{cite}
\usepackage{times}
\usepackage{graphicx}
\usepackage{amsmath}
\usepackage{algorithm}
\usepackage[noend]{algorithmic}
\usepackage{stfloats}
\usepackage{caption}
\usepackage{color}
\usepackage{fancyvrb}
\usepackage{amsmath}
\usepackage{amssymb}
\usepackage{makeidx}
\usepackage{mdwmath}
\usepackage{mdwtab}
\usepackage{moreverb}
\usepackage{fancyvrb}
\usepackage{etoolbox}
\usepackage{mathtools}
\usepackage{mathptmx}
\usepackage{multirow}
\usepackage{color}
\usepackage{url}
\usepackage{tabularx}
\usepackage{enumitem}
\usepackage[bookmarks=false]{hyperref}
\usepackage{footnote}

\makesavenoteenv{figure}

\usepackage{titlesec}
\titlespacing\section{0pt}{12pt plus 4pt minus 2pt}{2pt plus 2pt minus 2pt}
\titlespacing\subsection{0pt}{12pt plus 4pt minus 2pt}{2pt plus 2pt minus 2pt}

\usepackage{tikz}

\usepackage[nolist]{acronym}

\begin{acronym}[BEREC]
\acro{10G-EPON}{10 Gigabit EPON}
\acro{3GPP}{3rd Generation Partnership Project}
\acro{5G}{Fifth Generation}
\acro{6G}{Sixth Generation}
\acro{A-CPI}{A Controller Plane Interface}
\acro{A-RoF}{Analog Radio-over-Fiber}
\acro{ABNO}{Application-Based Network Operations}
\acro{ADC}{Analog-to-Digital Converter}
\acro{AE}{Allocative Efficiency}
\acro{amc}{adaptive modulation and coding}
\acro{AON}{Active Optical Network}
\acro{AI}{Artificial Intelligence}
\acro{ML}{Machine Learning}
\acro{API}{Application Programming Interface}
\acro{AWG}{Arrayed Waveguide Grating}
\acro{B2B}{Business to Business}
\acro{B2C}{Business to Consumer}
\acro{B-RAS}{Broadband Remote Access Servers}
\acro{BaaS}{Blockchain as a Service}
\acro{BB}{Budget Balance}
\acro{BBF}{BroadBand Forum}
\acro{BBU}{Base Band Unit}
\acro{BER}{Bit Error Rate}
\acro{BEREC}{Body of European Regulators for Electronic Communications}
\acro{BF}{beamforming}
\acro{BFT}{Byzentine fault tolerant}
\acro{BGP}{Border Gateway Protocol}
\acro{BMap}{Bandwidth Map}
\acro{BPM}{Business Process Management}
\acro{BS}{Base Station}
\acro{BSC}{Base Station Controller}
\acro{BtB}{Back to Back}
\acro{BW}{Bandwidth}
\acro{C2C}{Consumer to Consumer}
\acro{C-RAN}{Cloud Radio Access Network}
\acro{C-RoFN}{Cloud-based Radio over optical Fiber Network}
\acro{CA}{Certificate Authority}
\acro{CapEx}{Capital Expenditure}
\acro{CDMA}{code division multiple access}
\acro{CDR}{Clock Data Recovery}
\acro{CFO}{carrier frequency offset}
\acro{CIR}{Committed Information Rate}
\acro{CO}{Central Office}
\acro{CoMP}{Coordinated Multipoint}
\acro{COP}{Control Orchestration Protocol}
\acro{CORD}{Central Office Rearchitected as a Data Centre}
\acro{CPE}{Customer Premises Equipment}
\acro{CPRI}{Common Public Radio Interface}
\acro{CPU}{Central Processing Unit}
\acro{CMC}{Conflict Mitigation Controller}
\acro{CMS}{Conflict Management System}
\acro{CQI}{channel quality indicator}
\acro{CR}{cognitive radio}
\acro{CS}{Central Station}
\acro{CSI}{channel state information}
\acro{D2D}{Device to Device}
\acro{D-CPI}{D Controller Plane Interface}
\acro{D-RoF}{Digital Radio-over-Fiber}
\acro{DAC}{Digital-to-Analog Converter}
\acro{DAS}{distributed antenna system}
\acro{DAS}{Distributed Antenna Systems}
\acro{DBA}{Dynamic Bandwidth Allocation}
\acro{DBRu}{Dynamic Bandwidth Report upstream}
\acro{DC}{Direct Current}
\acro{DL}{Downlink}
\acro{DLT}{Distributed Ledger Technology}
\acro{DMT}{Discrete Multitone}
\acro{DNS}{Domain Name Server}
\acro{DPDK}{Data Plane Development Kit}
\acro{DSB}{Double Side Band}
\acro{DSL}{Digital Subscriber Line}
\acro{DSLAM}{Digital Subscriber Line Access Multiplexer}
\acro{DSP}{Digital Signal Processing}
\acro{DSS}{Distributed Synchronization Service}
\acro{DU}{Distributed Unit}
\acro{DUE}{D2D user equipment}
\acro{DWDM}{Dense Wave Division Multiplexing}
\acro{E-CORD}{Enterprise CORD}
\acro{EAL}{Environment Abstraction Layer}
\acro{EDF}{Erbium-Doped Fiber}
\acro{EFM}{Ethernet in the First Mile}
\acro{EMBS}{Elastic Mobile Broadband Service}
\acro{EON}{Elastic Optical Network}
\acro{EPC}{Evolved Packet Core}
\acro{EPON}{Ethernet Passive Optical Network}
\acro{eMBB}{enhanced Mobile Broadband}
\acro{ETSI}{European Telecommunications Standards Institute}
\acro{EVM}{Error Vector Magnitude}
\acro{FANS}{Fixed Access Network Sharing}
\acro{FASA}{Flexible Access System Architecture}
\acro{FBMC}{filter bank multi-carrier}
\acro{FBMC}{Filterbank Multicarrier}
\acro{FCC}{Federal Communications Commission}
\acro{FDD}{Frequency Division Duplex}
\acro{FDMA}{frequency-division multiple access}
\acro{FEC}{Forward Error Correction}
\acro{FFR}{Fractional Frequency Reuse}
\acro{FFT}{Fast Fourier Transform}
\acro{FPGA}{Field-Programmable Gate Array}
\acro{FSO}{Free-Space-Optics}
\acro{FSR}{Free Spectral Range}
\acro{FTN}{faster-than-nyquist}
\acro{FTTC}{Fiber-to-the-Curb}
\acro{FTTH}{Fiber-to-the-Home}
\acro{FTTx}{Fiber-to-the-x}
\acro{G.Fast}{Fast Access to Subscriber Terminals}
\acro{GEM}{G-PON encapsulation method}
\acro{GFDM}{Generalized Frequency Division Multiplexing}
\acro{GMPLS}{Generalized Multi-Protocol Label Switching}
\acro{GPON}{Gigabit Passive Optical Network}
\acro{GPP}{General Purpose Processor}
\acro{GPRS}{General Packet Radio Service}
\acro{GSM}{Global System for Mobile Communications}
\acro{GTP}{GPRS Tunneling Protocol}
\acro{GUI}{Graphical User Interface}
\acro{HA}{Hardware Accelerator}
\acro{HARQ}{Hybrid-Automatic Repeat Request}
\acro{HD}{half duplex}
\acro{HetNet}{heterogeneous networks}
\acro{IaaS}{Infrastructure as a Service}
\acro{I-CPI}{I Controller Plane Interface}
\acro{I2RS}{Interface 2 Routing System}
\acro{IA}{Interference Alignment}
\acro{IAM}{Identity and Access Management}
\acro{IBFD}{in-band full duplex}
\acro{IC}{Incentive Compatibility}
\acro{ICIC}{inter-cell interference coordination}
\acro{ICT}{information and communications technology}
\acro{IEEE}{Institute of Electrical and Electronics Engineers}
\acro{IETF}{Internet Engineering Task Force}
\acro{IF}{Intermediate Frequency}
\acro{IFFT}{inverse FFT}
\acro{ITS}{Intelligent Transport Systems}
\acro{InP}{Infrastructure Provider}
\acro{IoT}{Internet of Things}
\acro{IP}{Internet Protocol}
\acro{IQ}{In-phase/Quadrature}
\acro{IR}{Individual Rationality}
\acro{IRC}{Interference Rejection Combining}
\acro{ISI}{inter-symbol interference}
\acro{ISP}{Internet Service Provider}
\acro{BFT}{Byzantine-Fault-Tolerant}
\acro{IV}{Intelligent Vehicle}
\acro{ITU}{International Telecommunication Union}
\acro{KPI}{Key Performance Indicator}
\acro{KVM}{Kernel-based Virtual Machine}
\acro{L2}{Layer-2}
\acro{L3}{Layer-3}
\acro{LAN}{Local Area Network}
\acro{LO}{Local Oscillator}
\acro{LOS}{Line Of Sight}
\acro{LR-PON}{Long Reach PON}
\acro{LSP-DB}{Label Switched Path Database}
\acro{LTE-A}{Long Term Evolution Advanced}
\acro{LTE}{Long Term Evolution}
\acro{M-CORD}{Mobile CORD}
\acro{M-MIMO}{massive MIMO}
\acro{M2M}{machine-to-machine}
\acro{MAC}{Medium Access Control}
\acro{ME}{Merging Engine}
\acro{MIMO}{multiple input multiple output}
\acro{MME}{Mobility Management Entity}
\acro{mmWave}{millimeter wave}
\acro{MNO}{mobile network operator}
\acro{MPLS}{Multiprotocol Label Switching}
\acro{MRC}{Maximum Ratio Combining}
\acro{MSC}{Mobile Switching Centre}
\acro{MSP}{Membership Service Providers}
\acro{MSR}{Multi-Stratum Resources}
\acro{MTC}{machine type communication}
\acro{mMTC}{massive Machine Type Communications}
\acro{MVNO}{Mobile Virtual Network Operator}
\acro{MVNP}{mobile virtual network provider}
\acro{MZI}{Mach-Zehnder Interferometer}
\acro{MZM}{Mach-Zehnder Modulator}
\acro{NE}{Network Element}
\acro{NFV}{Network Function Virtualization}
\acro{NFVaaS}{Network Function Virtualization as a Service}
\acro{NG-PON2}{Next-Generation Passive Optical Network 2}
\acro{NGA}{Next Generation Access}
\acro{NLOS}{None Line Of Sight}
\acro{NMS}{Network Management System}
\acro{NRZ}{Non Return-to-Zero}
\acro{NTT}{Nippon Telegraph and Telephone}
\acro{OBPF}{Optical BandPass Filter}
\acro{OBSAI}{Open Base Station Architecture Initiative}
\acro{ODN}{Optical Distribution Network}
\acro{OFC}{Optical Frequency Comb}
\acro{OFDM}{orthogonal frequency-division multiplexing}
\acro{OFDMA}{orthogonal frequency-division multiple access}
\acro{OLO}{Other Licensed Operator}
\acro{OLT}{Optical Line Terminal}
\acro{ONF}{Open Networking Foundation}
\acro{ONU}{Optical Network Unit}
\acro{OOB}{out-of-band}
\acro{OOK}{On-off Keying}
\acro{OpenCord}{Central Office Re-Architected as a Data Center}
\acro{OpEx}{Operating Expenditure}
\acro{OSI}{Open Systems Interconnection}
\acro{OSS}{Operations Support Systems}
\acro{OTT}{Over-the-Top}
\acro{OXM}{OpenFlow Extensible Match}
\acro{P2MP}{Ethernet over point-to-multipoint }
\acro{P2P}{Point-to-Point }
\acro{PAM}{Pulse Amplitude Modulation}
\acro{PAPR}{peak-to-average power rating}
\acro{PAPR}{Peak-to-Average Power Ratio}
\acro{PBMA}{Priority Based Merging Algorithm}
\acro{PCE}{Path Computation Elements}
\acro{PCEP}{Path Computation Element Protocol}
\acro{PCF}{Photonic Crystal Fiber}
\acro{PD}{Photodiode}
\acro{PDCP}{RD Control Protocol}
\acro{PDF}{probability distribution function}
\acro{PGW}{Packet Gateway}
\acro{PHY}{physical Layer}
\acro{PIR}{Peak Information Rate}
\acro{PMD}{Polarization Division Multiplexing}
\acro{PON}{Passive Optical Network}
\acro{POTS}{Plain Old Telephone Service}
\acro{PoET}{Proof of Elapsed Time}
\acro{PoW}{Proof of Work}
\acro{PoS}{Proof of Stake}
\acro{PTP}{Precision Time Protocol}
\acro{PWM}{Pulse Width Modulation}
\acro{QAM}{Quadrature Amplitude Modulation}
\acro{QoE}{Quality of Experience}
\acro{QoS}{Quality of Service}
\acro{QPSK}{Quadrature Phase Shift Keying}
\acro{R-CORD}{Residential CORD}
\acro{RA}{Resource Allocation}
\acro{RAN}{Radio Access Network}
\acro{RAT}{Radio Access Technology}
\acro{RB}{resource block}
\acro{RFIC}{radio frequency integrated circuit}
\acro{RIC}{RAN Intelligent Controller}
\acro{RLC}{Radio Link Control}
\acro{RN}{Remote Node}
\acro{ROADM}{Reconfigurable Optical Add Drop Multiplexer}
\acro{RoF}{Radio-over-Fiber}
\acro{RRC}{Radio Resource Control}
\acro{RRH}{Remote Radio Head}
\acro{RRPH}{Remote Radio and PHY Head}
\acro{RRU}{Remote Radio Unit}
\acro{RSOA}{Reflective Semiconductor Optical Amplifier}
\acro{RU}{Remote Unit}
\acro{Rx}{receiver}
\acro{SC-FDMA}{single carrier frequency division multiple access}
\acro{SCM}{single carrier modulation}
\acro{SD-RAN}{Software Defined Radio Access Network}
\acro{SDAN}{Software Defined Access Network}
\acro{SDMA}{Semi-Distributed Mobility Anchoring}
\acro{SDN}{Software Defined Networking}
\acro{SDR}{Software Defined Radio}
\acro{SFBD}{Single Fiber Bi-Direction}
\acro{SGW}{Serving Gateway}
\acro{SI}{self-interference}
\acro{SIC}{self-interference cancellation}
\acro{SIMO}{Single Input Multiple Output}
\acro{SLA}{Service Level Agreement}
\acro{SMF}{Single Mode Fiber}
\acro{SNMP}{Simple Network Management Protocol}
\acro{SNR}{Signal-to-Noise Ratio}
\acro{SP}{Service Providers}
\acro{Split-PHY}{Split Physical Layer}
\acro{SRI-OV}{Single Root Input/Output Virtualization}
\acro{SU}{secondary user}
\acro{SUT}{System Under Test}
\acro{T-CONT}{Transmission Container}
\acro{TC}{Transmission Convergence}
\acro{TCO}{Total Cost of Ownership}
\acro{TD-LTE}{Time Division LTE}
\acro{TDD}{Time Division Duplex}
\acro{TDM}{Time Division Multiplexing}
\acro{TDMA}{Time Division Multiple Access}
\acro{TED}{Traffic Engineering Database}
\acro{TEID}{Tunnel endpoint identifier}
\acro{TLS}{Transport Layer Security}
\acro{TPS}{Transactions per Second}
\acro{TTI}{transmission time interval}
\acro{TWDM}{Time and Wavelength Division Multiplexing}
\acro{Tx}{transmitter}
\acro{UD-CRAN}{Ultra-Dense Cloud Radio Access Network}
\acro{UDP}{User Datagram Protocol}
\acro{UE}{User Equipment}
\acro{UF-OFDM}{Universally Filtered OFDM}
\acro{UFMC}{Universally Filtered Multicarrier}
\acro{UL}{Uplink}
\acro{UMTS}{Universal Mobile Telecommunications System}
\acro{URLLC}{Ultra-Reliable Low-Latency Communication}
\acro{USRP}{Universal Software Radio Peripheral}
\acro{V2I}{Vehicle to Infrastructure}
\acro{V2V}{Vehicle to Vehicle}
\acro{vBBU}{virtualized BBU}
\acro{vBMap}{Virtual Bandwidth Map}
\acro{vBS}{virtual Base station}
\acro{VCG}{Vickery-Clarke-Groves}
\acro{vCPE}{virtual CPE}
\acro{vCPU}{Virtual Central Processing Unit}
\acro{vDBA}{virtual DBA}
\acro{VDSL2}{Very-high-bit-rate digital subscriber line 2}
\acro{VLAN}{Virtual Local Area Network}
\acro{VM}{Virtual Machine}
\acro{VNF}{Virtual Network Function}
\acro{VNI}{visual networking index}
\acro{VNO}{Virtual Network Operator}
\acro{VNTM}{Virtual Network Topology Manager}
\acro{vOLT}{virtual OLT}
\acro{VPE}{virtual Provider Edge}
\acro{VPN}{Virtual Private Network}
\acro{V2X}{Vehicle to Everything}
\acro{VTN}{Virtual Tenant Network}
\acro{VULA}{Virtual Unblunded Local Access}
\acro{WAN}{Wide Area Network}
\acro{WAP}{Wireless Access Point}
\acro{WCDMA}{wide-band code division multiple access}
\acro{WDM-PON}{Wavelength Division Multiplexing - Passive Optical Network}
\acro{WDM}{Wavelength Division Multiplexing}
\acro{WiMAX}{Worldwide Interoperability for Microwave Access}
\acro{WRPR}{Wired-to-RF Power Ratio}
\acro{XG-PON}{10 Gigabit PON}
\acro{XGEM}{XG-PON encapsulation method}
\acro{XOS}{XaaS Operating System}

\acro{RIC}{RAN Intelligent Controller}
\acro{Near-RT-RIC}{Near Real Time RAN Intelligent Controller}
\acro{Non-RT-RIC}{Non Real Time RAN Intelligent Controller}
\acro{xApp}{Extended Application}
\acro{rApp}{Remote Application}
\acro{CMS}{Conflict Mitigation System}
\acro{EG}{Eisenberg-Galle}
\acro{NSWF}{Nash's Social Welfare Function} 
\acro{CMC}{Conflict Mitigation Controller}
\acro{CDC}{Conflict Detection Controller}
\acro{PMon}{Performance Monitoring}
\acro{RCP}{Recently Changed Parameter}
\acro{PGD}{Parameter Group Definition}
\acro{RCPG}{Recently Changed Parameter Group}
\acro{DCKD}{Decision Correlated with KPI Degradation} 
\acro{KDO}{KPI Degradation Occurrences}
\acro{MNO}{Mobile Network Operator}
\acro{SDL}{Shared Data Layer}
\acro{SON}{Self Organising Network}
\acro{SF}{SON Function}
\acro{AM}{Arithmetic Mean}
\acro{PKR}{Parameter and KPI Ranges}
\acro{KPI}{Key Performance Indicator}
\acro{ICP}{Input Control Parameter}
\acro{RF}{Radio Frequency}
\acro{C-RAN}{Cloud Radio Access Network}
\acro{MARL}{Multi Agent Reinforcement Learning}
\acro{CCO}{Capacity and Coverage Optimization}
\acro{ES}{Energy Saving}
\acro{MRO}{Mobility Robustness Optimization}
\acro{MLB}{Mobility Load Balancing}
\acro{V-RAN}{Virtualized Radio Access Network}
\acro{O-RAN}{Open Radio Access Network}
\acro{D-RAN}{Distributed Radio Access Network}
\acro{5GB}{5G and Beyond}
\acro{O-RU}{Open Radio Unit}
\acro{O-DU}{Open Distribution Unit}
\acro{O-CU}{Open Centralised Unit}
\acro{O-CU-CP}{Open Centralised Unit - Control Plane}
\acro{O-CU-UP}{Open Centralised Unit - User Plane}
\acro{D-SON}{Distributed SON}
\acro{C-SON}{Centralised SON}
\acro{CIO}{Cell Individual Offset}
\acro{CBR}{Call Block Ratio}
\acro{HOR}{Handover Ratio}
\acro{CAN}{Cognitive Autonomous Network}
\acro{CF}{Cognitive Function}
\acro{QACM}{QoS-Aware Conflict Mitigation}
\acro{ANN}{Artificial Neural Network}
\acro{PR}{Polynomial Regression}
\acro{MSE}{Mean Squared Error}
\acro{EVS}{Explained Variance Score}
\acro{CS}{Conflict Supervision}
\acro{TXP}{Tx Power}
\acro{TTT}{Time-To-Trigger}
\acro{CIO}{Cell Individual Offset}
\acro{NL}{Neighbour List}
\acro{RET}{Radio Electrical Tilt}
\acro{HYS}{Handover Hysteresis}
\acro{QACMP}{QACM Priority}
\acro{DRL}{Deep Reinforcement Learning}
\acro{gNB}{Next-generation NodeB}
\acro{RSRP}{Reference Signal Received Power}
\end{acronym}

\begin{document}

\title{
    \vspace{-0cm} 
    \begin{tikzpicture}[remember picture, overlay]
        \node[anchor=north, yshift=-0.5cm] at (current page.north) {\fbox{\parbox{\textwidth}{\centering\small {\color{red}This version of the paper has been accepted for publication to IEEE INFOCOM and may be removed upon request or copyright issues.}}}};
    \end{tikzpicture}
    \vspace{0cm} 
    xApp-Level Conflict Mitigation in O-RAN, a Mobility Driven Energy Saving Case

}



\author{\IEEEauthorblockN{Abdul Wadud\IEEEauthorrefmark{1}\IEEEauthorrefmark{2},~\IEEEmembership{Graduate Student Member,~IEEE}
Fatemeh Golpayegani\IEEEauthorrefmark{1},~\IEEEmembership{Senior Member,~IEEE} and} 
\IEEEauthorblockN{Nima Afraz \IEEEauthorrefmark{1},~\IEEEmembership{Senior Member,~IEEE}}
\IEEEauthorblockA{\IEEEauthorrefmark{1}School of Computer Science,
University College Dublin, Ireland}
\IEEEauthorblockA{\IEEEauthorrefmark{2}Bangladesh Institute of Governance and Management, Dhaka, Bangladesh}
\thanks{Corresponding author: Abdul Wadud (email: abdul.wadud@ucdconnect.ie).}}

%



\IEEEtitleabstractindextext{%
\begin{abstract}
This paper addresses the  challenges of conflict detection and mitigation in Open \ac{RAN} where multiple vendors co-operate the same \ac{RAN} and share access to control parameters and conflicts between \acp{xApp} can arise that affect network performance and stability due to the disaggregated nature of Open \ac{RAN}. This work provides a detailed theoretical framework of \ac{xApp}-level conflicts, i.e., direct, indirect, and implicit conflicts. Leveraging conflict graphs, we further highlight how conflicts impact \acp{KPI} and propose a low-latency rule-based strategy for conflict detection using \acp{SLA} and \ac{QoS} thresholds. We evaluate the effectiveness of several mitigation strategies in a simulated environment with \ac{MRO} and \ac{ES} \acp{xApp} and present experimental results showing comparisons among these strategies. The findings of this research provide significant insights for enhancing O-RAN deployments with flexible and efficient conflict management.
\end{abstract}

\begin{IEEEkeywords}
O-RAN, Conflict, SLA, QoS, MRO, ES, xApp, and KPI.
\end{IEEEkeywords}}

\IEEEoverridecommandlockouts
\IEEEpubid{\makebox[\columnwidth]{978-1-XXXX-XXXX-X/XX/\$31.00~\copyright~2025 IEEE} \hfill}
\IEEEpubidadjcol

\maketitle
\thispagestyle{plain}
\pagestyle{plain}

\IEEEdisplaynontitleabstractindextext

%
\IEEEpeerreviewmaketitle

\section{Introduction}
\label{sec:intro}
\IEEEPARstart{T}{he} introduction of Open \ac{RAN} architecture has promised enhanced flexibility and interoperability in the telecommunications industry. However, this disaggregated approach of integrating components from multiple vendors presents a significant challenge of conflict management. While the traditional \ac{RAN} relied on single-vendor solutions that come with in-house vendor specific conflict resolution strategies, Open \ac{RAN}'s multi-vendor nature needs a robust conflict mitigation strategy to cope with conflicts between different vendors' components. Conflicts arise when components compete for shared resources or set conflicting configurations that control the same network environment. This can lead to performance degradation, instability, and security vulnerabilities in the network. Therefore, handling these conflicts has become a priority and a major obstacle to the widespread deployment of Open \ac{RAN}.

A key challenge is that the conflict definitions provided by the O-RAN Alliance are not comprehensive, leading to various interpretations of conflicts. Some research contradicts the basic conflict definitions established by the O-RAN Alliance in \cite{ric_oran_alliance}. For instance, in \cite{del2024pacifista, zolghadr2024learning}, the definition of implicit conflict aligns with what is typically described as an indirect conflict—where different parameters belonging to separate \acp{xApp} influence the same \ac{KPI}. To provide context, the O-RAN Alliance defines that Antenna Tilt (or \ac{RET}), associated with \ac{MLB}/\ac{CCO} xApps, affects the handover boundary, while the measurement offset related to \ac{MRO} \ac{xApp} has a similar impact. This results in handover-related \acp{KPI} being influenced by parameters from distinct \acp{xApp}, leading to what is described as an indirect conflict. However, in these studies, the definition of indirect conflict has been altered and labeled as implicit conflict that makes their entire detection and mitigation strategy flawed. Similarly, their definition of indirect conflict describes a parameter-to-parameter indirect relationship, which does not align with the O-RAN Alliance’s definition of parameter-to-KPI relationship. Therefore, this paper provides a comprehensive guide to conflict definitions in Open RAN, ensuring alignment with the O-RAN Alliance’s framework.

While significant progress has been made in understanding and mitigating conflicts in Open RAN, several challenges remain: 
\begin{itemize}
    \item Knowledge Gap: As Open \ac{RAN} is a new concept, and there is limited knowledge and misconceptions about conflicts within it, therefore, this paper provides a comprehensive discussion on various types of conflicts in Open \ac{RAN}. 
    \item Indirect vs Implicit Conflict: The subtle difference between indirect and implicit conflicts makes reliably detecting and evaluating these potential issues challenging. Developing a low-latency detection mechanism is crucial to ensure the reliability and stability of Open \ac{RAN} deployments.
    \item Dynamic Detection and Mitigation: Existing solutions often focus on static conflict resolution. However, the dynamic nature of network traffic and user demands necessitates adaptive and real-time conflict detection and mitigation strategies.
    \item Scalability and Complexity: As Open \ac{RAN} networks expand in size and complexity, managing conflicts between numerous \acp{xApp} and \acp{rApp} from diverse vendors will become increasingly challenging. Scalable and efficient conflict detection and mitigation mechanisms are essential for the large-scale adoption of Open RAN.
\end{itemize}

\noindent In light of these challenges, this research focuses on:
\begin{itemize}
    \item Studying the definition of conflicts based on state-of-art and O-RAN Alliance guideline in a comprehensive way to clear misconceptions. 
    \item Providing a detailed theoretical framework to explain each type of \ac{xApp} conflict that helps simplifying conflict detection and mitigation.
    \item Utilizing conflict graphs for more clear representation and better understanding of potential conflict.
    \item Identifying differences between conflict types systematically and developing a low-latency conflict detection framework adaptable to evolving network conditions.
    \item Reviewing and analyzing the performance of state-of-the-art mitigation strategies.
\end{itemize}

The structure of the paper is as follows: Section~\ref{sec:conflict_cat} provides a comprehensive overview of conflicts in Open \ac{RAN}, Section~\ref{sec:xApp_conflict} discusses the system model and discusses \ac{xApp} conflicts with a theoretical model, Section~\ref{sec:conflict_detection} presents a discussion on the rule-based conflict detection method and \ac{QACM} mitigation strategy, Section~\ref{sec:simRes} discusses simulation experiments and results with \ac{ES} and \ac{MRO} \acp{xApp}, Section~\ref{sec:limitF} presents the limitations of this work indicating future research directions, and the paper concludes in Section~\ref{sec:conclusion}.

\section{Background Study and State-of-the-Art}
\label{sec:conflict_cat}
This section provides an initial understanding of different types of conflicts in Open \ac{RAN} based on the state-of-the-art. Conflicts are categorized into vertical and horizontal conflicts, which are further classified into intra-component and inter-component conflicts \cite{adamczyk2023conflict, wadud2023conflict, wadud2024qacm}.

\textbf{Vertical conflicts} involve components from different hierarchical levels, such as a conflict between a \ac{Near-RT-RIC} and a Non-RT RIC \cite{wadud2023conflict, wadud2024qacm}. \textbf{Horizontal conflicts} occur between components at the same hierarchical level, for instance, between two \acp{xApp} (applications for near-real-time control) within a \ac{Near-RT-RIC} \cite{wadud2023conflict}. It is further classified into intra-component and inter-component conflicts:
    \begin{itemize}
        \item Intra-component conflicts occur within a component, like conflicting configurations between \acp{xApp} in a \ac{Near-RT-RIC} \cite{wadud2023conflict, wadud2024qacm}. These are further classified into three types: 
        \begin{itemize}
            \item \textit{Direct Conflict:} These occur when two or more \acp{xApp} attempt to simultaneously control the same network parameter \cite{adamczyk2023conflict, erdol2024xapp, wadud2023conflict, del2024pacifista, ric_oran_alliance}. For instance, one \ac{xApp} might try to increase the transmission power for a specific cell, while another aims to decrease it for energy saving. Direct conflicts are generally easier to detect since the conflicting parameter is directly observable \cite{polese2022understanding, wadud2024qacm}. However, simply prioritizing one \ac{xApp}'s preference over another might not be the optimal solution. A more effective approach would involve finding a configuration that balances these conflicting objectives, potentially maximizing overall network utility \cite{wadud2023conflict}.

            \item \textit{Indirect Conflicts:} These arise when different \acp{xApp} control distinct parameters that ultimately influence the \ac{KPI} \cite{adamczyk2023conflict, ric_oran_alliance, adamczyk2023detection, wadud2023conflict, corici2024towards}. For example, \ac{MRO} \ac{xApp} might modify the cell individual offset (CIO) to adjust the handover success rate and reduce link failures, while \ac{MLB} adjusts antenna tilts to balance the traffic loads. Both actions indirectly impact the cell boundaries and handover decisions, potentially causing conflicts. Detecting indirect conflicts necessitates post-action analysis, observing how different \ac{xApp} actions collectively impact the relevant \ac{KPI}. This is more challenging than detecting direct conflicts as the relationship between parameter changes and \ac{KPI} impacts might not be immediately apparent \cite{wadud2024qacm}.

            \item \textit{Implicit Conflicts:} These present the most significant challenge for detection and mitigation \cite{adamczyk2023conflict, polese2022understanding}. Implicit conflicts occur when the actions of multiple \acp{xApp}, while individually aligned with their specific objectives, result in an undesirable overall network state \cite{erdol2024xapp, adamczyk2023detection}. This often involves subtle interactions between \acp{xApp} with different goals, making the source of the conflict difficult to pinpoint. For example, a SLA-assuring \ac{xApp} trying to maximize quality of service (QoS) for a group of users might interfere with another \ac{xApp} aiming to minimize handover rates, leading to unexpected performance degradation. Identifying and resolving implicit conflicts often requires sophisticated monitoring and analysis of network behavior over time \cite{adamczyk2023conflict}.
        \end{itemize}
        \item Inter-component conflicts involve components from different areas, such as conflicting decisions from \acp{xApp} in neighboring \ac{Near-RT-RIC}s \cite{wadud2024qacm, wadud2023conflict}.
    \end{itemize}
     
Addressing various conflicts requires a multi-pronged approach. Existing literature suggests several strategies: 

\subsubsection{Conflict Detection and Mitigation Frameworks:}
Adamczyk and Kliks \cite{adamczyk2023conflict} propose a conflict mitigation framework (CMF) within the \ac{Near-RT-RIC}. The framework focuses on detecting direct, indirect, and implicit conflicts between \acp{xApp}. Direct conflicts arise from consecutive, contradictory decisions affecting the same parameters. Indirect conflicts occur when \acp{xApp} impact shared \acp{KPI} through different parameters. Implicit conflicts involve \ac{xApp} actions aligning with individual objectives but contradicting overall network goals. A more detailed description of these conflicts are discussed in Section.~\ref{sec:xApp_conflict}. The framework in \cite{adamczyk2023conflict} utilizes a Conflict Mitigation (CM) component in the \ac{Near-RT-RIC} to identify and resolve these conflicts. However, current implementations primarily focus on specific \acp{xApp}, like Mobility Robustness Optimization (MRO) and Mobility Load Balancing (MLB). 

\subsubsection{Team Learning and Cooperative Approaches:}
Zhang et al. in \cite{zhang2022team} suggest a team learning algorithm based on Deep Q-learning to encourage cooperation between \acp{xApp}. By sharing information about their intended actions, \acp{xApp} can make more informed decisions that benefit the overall network performance. This approach demonstrates promising results, leading to improved throughput and reduced packet drop rates. However, scalability remains a concern, as the complexity increases with the number of cooperating \acp{xApp}.

\subsubsection{QoS-Aware Conflict Mitigation:}
Recognizing the importance of maintaining Quality of Service (QoS), our previous work \cite{wadud2024qacm} introduces the QACM (QoS-Aware Conflict Mitigation) method. QACM considers the QoS benchmarks of individual \acp{xApp} during conflict mitigation, ensuring their requirements are met. It utilizes game theory principles, specifically Nash’s Social Welfare Function (NSWF) and Eisenberg-Galle (EG) solutions, to find an optimal balance between conflicting parameters. Initial results indicate QACM's effectiveness in upholding QoS thresholds compared to benchmark methods. However, further research is needed to develop more sophisticated \ac{KPI} prediction models and test its practicality in real-world RAN deployments.

\section{System Model}
\label{sec:xApp_conflict}
We develop a theoretical framework for modeling intra-component conflicts, focusing on the interactions between different \acp{xApp} within the \ac{Near-RT-RIC}. This framework uses conflict graphs to represent potential conflict points and their characteristics. Let us consider, there are $n$ number of \acp{xApp} present in a \ac{Near-RT-RIC}, represented by $x_n \in X$. Additionally, there are $i$ numbers of \acp{ICP} and $j$ number of \acp{KPI}, each represented as $p_i \in P$ and $k_j \in K$, respectively. Each \ac{xApp} has a set of \acp{ICP} denoted by $I_{x_n}$ and a set of \acp{KPI} represented by $K_{x_n}$. Intra-component conflicts are of three types as stated above in Section~\ref{sec:conflict_cat}.
    
\begin{figure}[!ht]
 \centering
\includegraphics[scale=0.55]{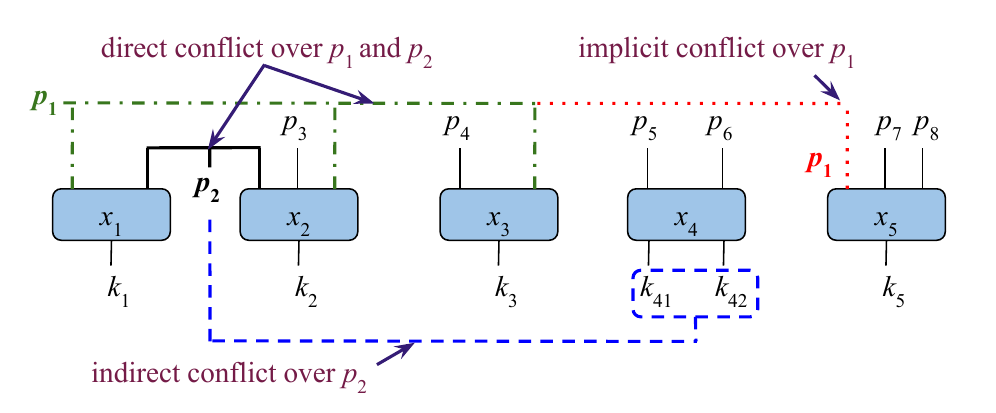}
	\vspace{-0.1in}
	\caption{An example model of conflict with five stochastic \acp{xApp} \cite{wadud2024qacm}.}
	\label{fig:con_example}
 \vspace{-0.1in}
\end{figure}

We consider an example model with five stochastic \acp{xApp} in Fig.~\ref{fig:con_example} to explain each of the three types of conflicts. We consider only five \acp{xApp}, because, our main goal is to cover all three types of conflicts and represent them in the most reader-friendly way in a limited space. 

\begin{figure}
    \centering
    \includegraphics[scale=0.3]{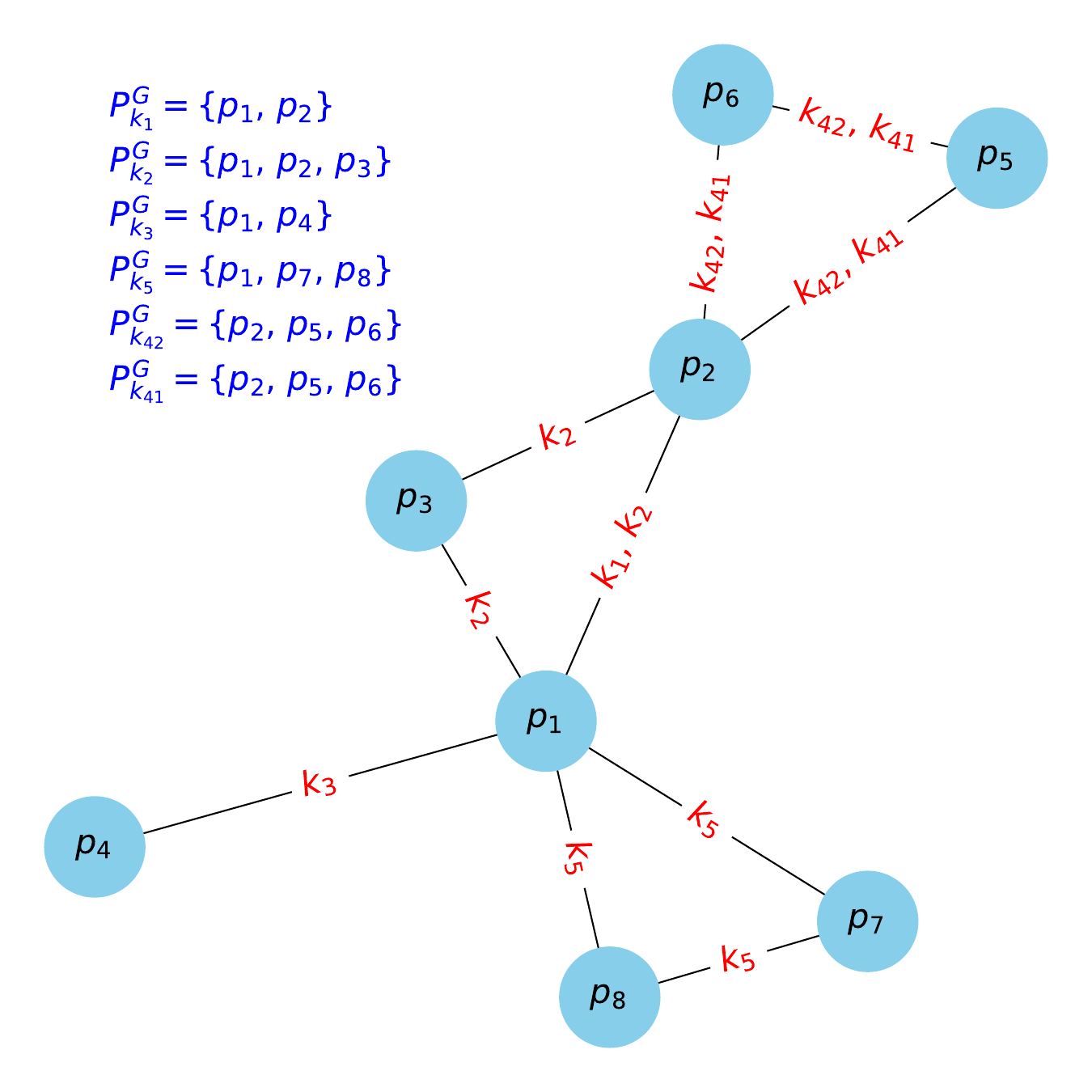}
    \caption{K-P Graph with \acp{KPI} for Parameters in Open RAN.}
    \label{fig:kp_graph}
\end{figure}

In Fig.~\ref{fig:con_example}, the set of input control parameters and \acp{KPI} for each \ac{xApp} is defined as follows: $I_{x_1} = \{p_1, p_2\}$, $I_{x_2} = \{p_1, p_2, p_3\}$, $I_{x_3} = \{p_1, p_4\}$, $I_{x_4} = \{p_5, p_6\}$, $I_{x_5} = \{p_7, p_8\}$, $K_{x_1} = \{k_1\}$, $K_{x_2} = \{k_2\}$, $K_{x_3} = \{k_3\}$, $K_{x_4} = \{k_{41}, k_{42}\}$, and $K_{x_5} = \{k_5\}$. The set of all \acp{xApp} considered in this study is:
$ X = \{x_1, x_2, x_3, x_4, x_5\}$. The set of \acp{ICP} and \acp{KPI} considered in this study is: $P = \{p_1, p_2, p_3, p_5, p_5, p_6, p_7\}$ and
$K = \{k_1, k_2, k_3, k_5, k_{41}, k_{42}\}$. The parameter group for each \ac{KPI} is formulated based on the control parameters that influence the \ac{KPI}. The Algorithm~\ref{algo:param_kpi_grouping} and K-P graph in Fig.~\ref{fig:kp_graph} helps in identifying these groups. For example: $P_{k_1}^G = \{p_1, p_2\}$, $P_{k_2}^G = \{p_1, p_2, p_3\}$, $P_{k_3}^G = \{p_1, p_4\}$, $P_{k_5}^G = \{p_7, p_8\}$, $P_{k_{41}}^G = \{p_2, p_5, p_6\}$, $P_{k_{42}}^G = \{p_2, p_5, p_6\}$. To understand how parameter group for each \ac{KPI} is formed, let us consider $k_1$. From the K-P graph in Fig.~\ref{fig:kp_graph}, we see that $p_1$ and $p_2$ are parameters that influence $k_1$. Therefore, the parameter group for $k_1$ is:
$P_{k_1}^G = \{p_1, p_2\}$.

\subsubsection{Direct Conflict}
\begin{figure}
    \centering
    \includegraphics[scale=0.3]{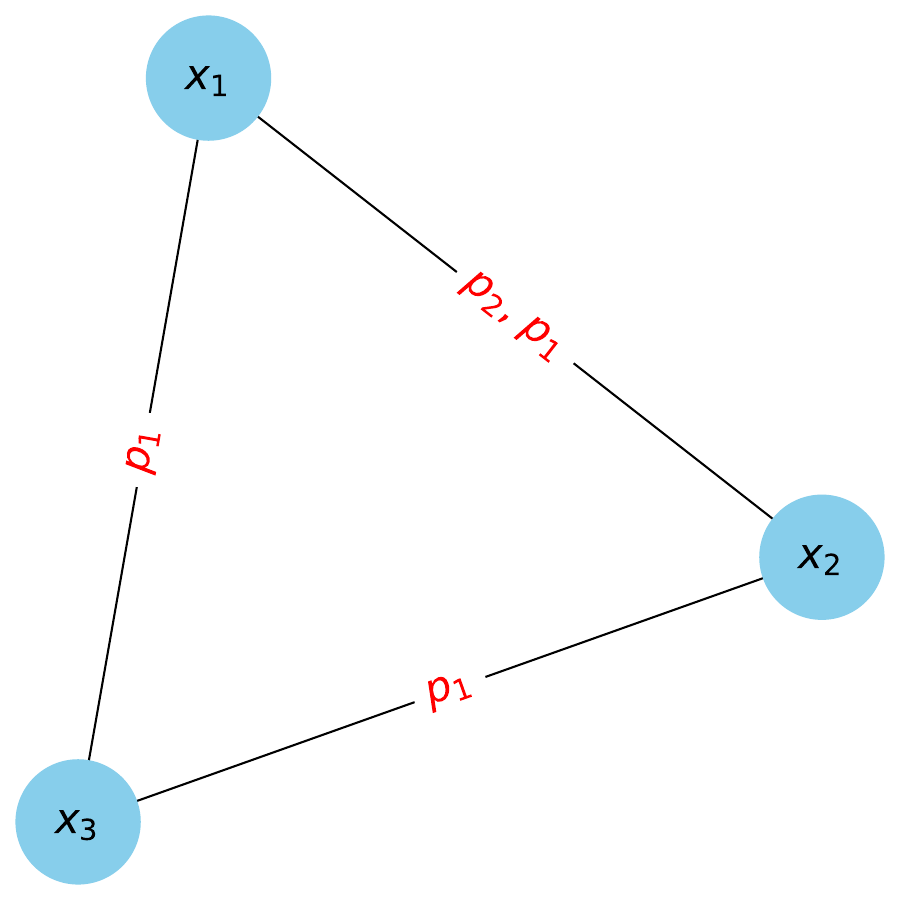}
    \caption{X-P Graph with \acp{ICP} for \acp{xApp} in Open RAN.}
    \label{fig:xp_graph}
\end{figure}

A direct conflict occurs when an intersection operation between the set of input control parameters of two or more \acp{xApp} results in a non-null set. This indicates that these \acp{xApp} have conflicting input parameters. Based on the X-P graph in Fig.~\ref{fig:xp_graph}, specific direct conflict cases can be identified as follows:

\begin{itemize}
    \item $I_{x_1} \cap I_{x_2} = \{p_1, p_2\}$: There is a direct conflict between $x_1$ and $x_2$ over the parameters $p_1$ and $p_2$.
    \item $I_{x_1} \cap I_{x_3} = \{p_1\}$: There is a direct conflict between $x_1$ and $x_3$ over the parameter $p_1$.
    \item $I_{x_2} \cap I_{x_3} = \{p_1\}$: There is a direct conflict between $x_2$ and $x_3$ over the parameter $p_1$.
\end{itemize}

\subsubsection{Indirect Conflict}
An indirect conflict occurs when a group of input control parameters, belonging to different \acp{xApp}, are associated with a common \ac{KPI}. This type of conflict can be identified using the K-P graph. For instance:
\begin{itemize}
    \item $P_{k_{41}}^G = \{p_2, p_5, p_6\}$ and $P_{k_{42}}^G = \{p_2, p_5, p_6\}$: Here, $p_2 \notin I_{x_4}$, but it still affects $k_{41}$ and $k_{42}$. Since $p_2 \in I_{x_1}$ and $I_{x_2}$, we can say there is an indirect conflict of these \acp{xApp} with $x_4$ over $p_2$.
\end{itemize}

\subsubsection{Implicit Conflict}
An implicit conflict is identified when a \ac{KPI} is affected by a parameter not originally considered part of its control parameter group. For example:
\begin{itemize}
    \item $P_{k_5}^G = \{p_7, p_8\}$: This parameter group indicates that only modifying $p_7$ and $p_8$ should affect $k_5$. However, if during the RAN operation, $k_5$ gets affected by changing $p_1$, then we can say there is an implicit conflict between $x_1, x_2, x_3$ and $x_5$ over $p_1$ since $p_1 \in I_{x_1}$, $I_{x_2}$, and $I_{x_3}$. Including $p_1$ inside $P_{k_5}^G$ changes it to $P_{k_5}^G = \{p_1, p_7, p_8\}$, making it an indirect conflict. Thus, an implicit conflict is a vague form of indirect conflict that can only be detected during RAN operation but can be modeled as an indirect conflict once detected.
\end{itemize}

\begin{table*}[ht]
\centering
\caption{Comparison of Different \acp{xApp} Conflict Types in Open RAN}
\label{tab:conCompare}
\resizebox{\textwidth}{!}{%
\begin{tabular}{|c|c|c|c|}
\hline
\textbf{Conflict Type} & \textbf{Definition} & \textbf{Mathematical Notation} & \textbf{Example Conflict} \\ \hline
\textbf{Direct Conflict} & Occurs when two or more \acp{xApp} share common input parameters & $I_{x_i} \cap I_{x_j} \neq \emptyset$ & $I_{x_1} \cap I_{x_2} = \{p_1, p_2\}$ \\ \hline
\textbf{Indirect Conflict} & Occurs when different \acp{xApp}' input parameters influence the same \ac{KPI} & $P_{k_i}^G = \{p_x, p_y\}, P_{k_j}^G = \{p_x, p_y\}$ & $P_{k_{41}}^G = \{p_2, p_5, p_6\}$ \\ \hline
\textbf{Implicit Conflict} & Occurs when a \ac{KPI} is influenced by parameters outside its control group & $P_{k_i}^G = \{p_x, p_y\} \Rightarrow p_z \text{ influences } k_i$ & $P_{k_5}^G = \{p_7, p_8\} \text{ but } p_1 \text{ affects } k_5$ \\ \hline
\end{tabular}%
}
\end{table*}

\section{Conflict Detection and Mitigation}
\label{sec:conflict_detection}

Detecting conflicts is one of the most challenging aspects of conflict management in the \ac{Near-RT-RIC}. It requires a deep understanding of each type of conflict and their distinctions. The comprehensive theoretical analysis presented in Section~\ref{sec:xApp_conflict} and the comparison in Table~\ref{tab:conCompare} provide the foundation for a rule-based, low-latency conflict detection mechanism. \ac{xApp}-level conflicts are typically parametric and can often be detected by tracing \acp{ICP} and monitoring \ac{KPI} degradation events. Some \acp{KPI} are critical to maintain as per \ac{SLA} agreements, while others are less so. 

In the conflict management system proposed in \cite{wadud2023conflict, wadud2024qacm}, the \ac{MNO} is responsible for defining the important \acp{KPI} that must be maintained at predefined thresholds. This paper adopts the \ac{CMS} presented in our previous work \cite{wadud2023conflict, wadud2024qacm}, which includes separate \ac{CDC} and \ac{CMC} components within the \ac{Near-RT-RIC}. The \ac{CDC} communicates back and forth with the \ac{Near-RT-RIC} database while performing detection tasks. 

The \ac{Near-RT-RIC} database consists of several components, including \ac{RCP}, \ac{PGD}, \ac{RCPG}, \ac{PKR}, \ac{DCKD}, and \ac{KDO}. When a \ac{xApp} is deployed in the \ac{Near-RT-RIC}, the \ac{MNO} is expected to provide the xApp details, including their \acp{ICP} ($I_{x_n}$) and \acp{KPI} ($K_{x_n}$). The \ac{PGD} component of the database is equipped with Algorithm~\ref{algo:param_kpi_grouping}, which estimates the $P_k^G$ for all $k_j \in K_{x_n}$. The \ac{KDO} component is responsible for storing and tracking \ac{KPI} degradation events. If any \ac{SLA}-sensitive \ac{KPI} falls below the predefined threshold, it alerts the \ac{CDC} of a possible conflict, along with the timestamp ($t_{clock}$) indicating when the event occurred.

\begin{figure}
\centering
\includegraphics[scale=0.5]{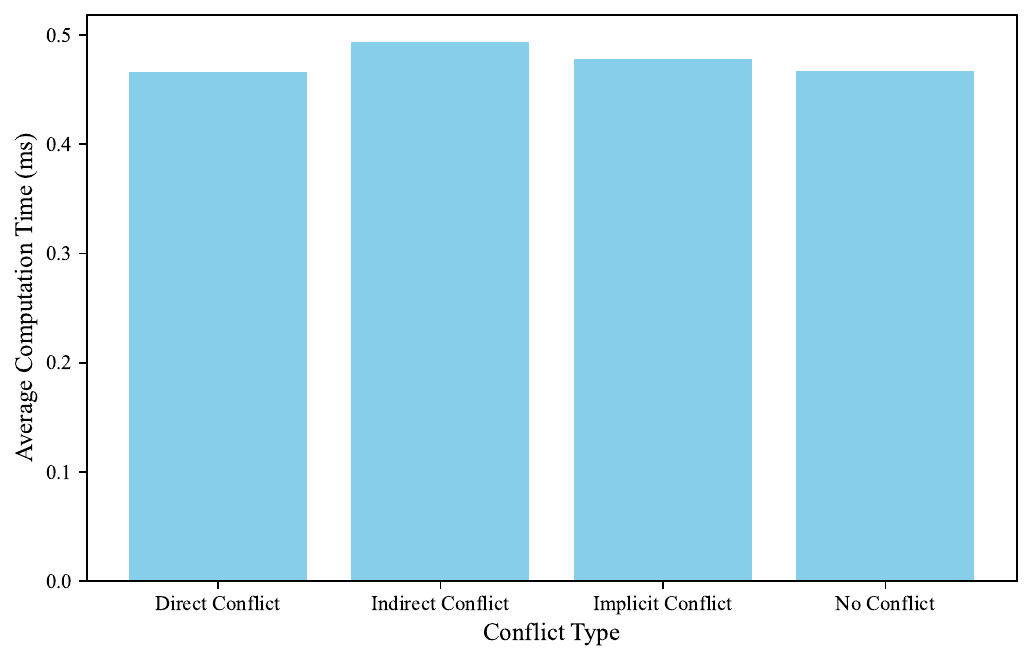}
\caption{Average Computation Time for Conflict Detection.}
\label{fig:conDetectPerf}
\end{figure}

\begin{figure}
\centering
\includegraphics[scale=0.32]{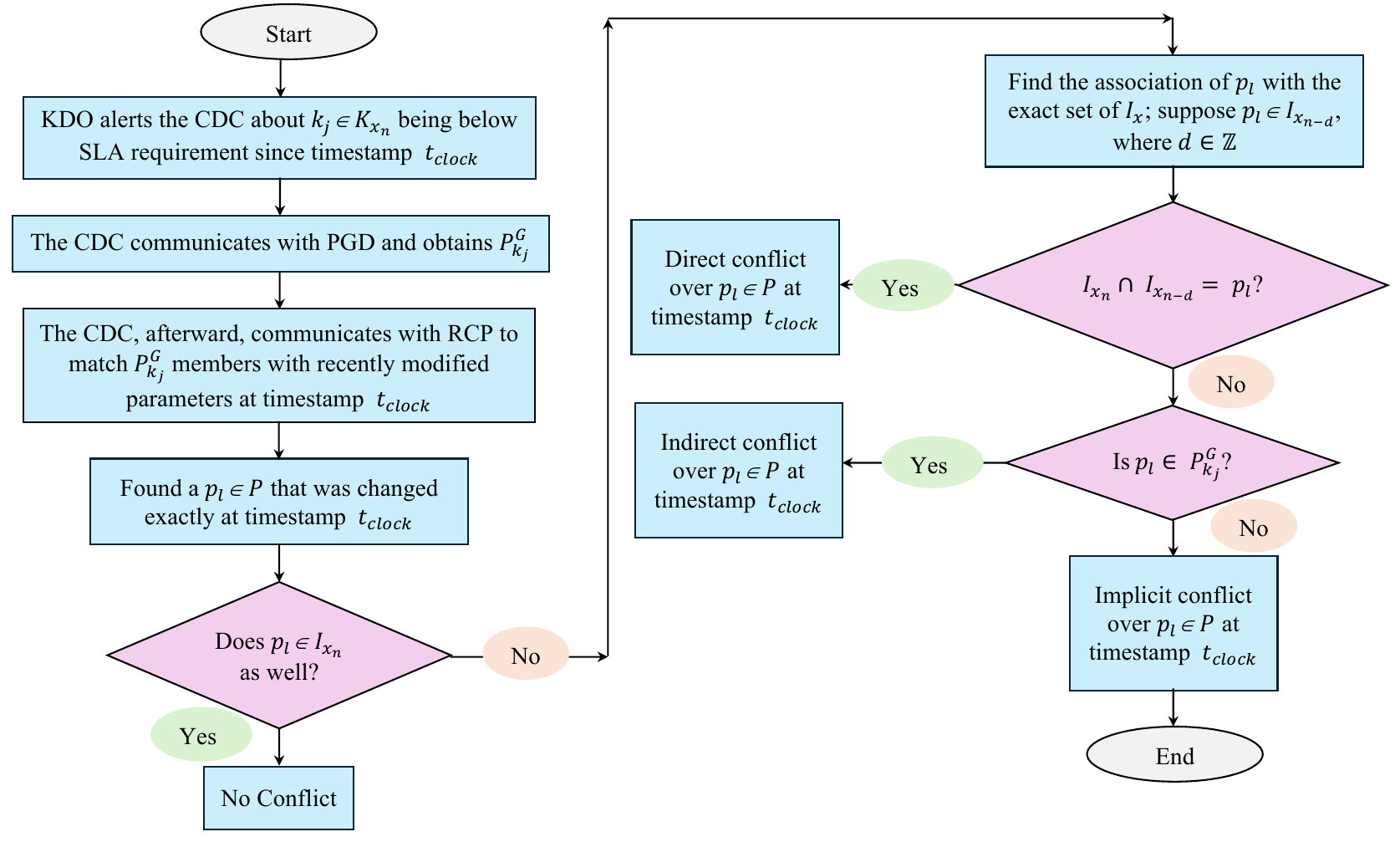}
\caption{Rule-based conflict detection mechanism.}
\label{fig:conDetect}
\end{figure}

The \ac{RCP} component tracks any parameter changes, aiding the \ac{CDC} in identifying the exact parameter responsible for the conflict at $t_{clock}$. The step-by-step conflict detection process is illustrated in Fig.~\ref{fig:conDetect}, which explains how a rule-based detection mechanism can identify specific types of conflicts. To analyze the performance of the proposed rule-based detection method, we adopted the stochastic xApp model presented in \cite{wadud2023conflict, wadud2024qacm} with the same configuration illustrated in Fig.~\ref{fig:con_example}. The \ac{CDC} is triggered only when a \ac{KPI} violation occurs; otherwise, it is considered a no-conflict state. When the instructing \ac{xApp} that modifies the conflicting parameter is the same as the \ac{xApp} associated with the degraded \ac{KPI}, we classify it as 'no conflict.' If the instructing \ac{xApp} and the \ac{xApp} associated with the degraded \ac{KPI} are different but share the conflicting parameter, it is detected as a 'direct conflict.' For an 'indirect conflict,' we check the parameter group of the degraded \ac{KPI}; if the conflicting parameter belongs to this group and the case has not been classified as 'no conflict' or 'direct conflict,' we consider it an 'indirect conflict.' Finally, if none of the above three conditions are met, the conflict is classified as 'implicit conflict.' The rule-based detection method shows an average latency of 0.4 to 0.5 milliseconds for detecting each type of conflict as shown in Fig.~\ref{fig:conDetectPerf} with a 100 percent accuracy.

\begin{algorithm}[!ht]
\caption{Parameter Grouping Based on KPIs for xApps}
\begin{algorithmic}[1]
\REQUIRE A set of xApps with their ICPs $I_{x_n}$ and KPIs $K_{x_n}$.
\ENSURE Parameter group $P_k^G$ for $k_j$.
\STATE Initialize a dictionary $param\_to\_kpis$ to map each ICP to its associated KPIs.
\STATE Initialize a dictionary $P_k^G$ to map each KPI to its associated ICPs.
\FOR{each xApp $x_n$}
    \FOR{each ICP $p_i \in I_{x_n}$}
        \FOR{each KPI $k_j \in K_{x_n}$}
            \STATE Add $k_j$ to the list of KPIs for $p_i$ in $param\_to\_kpis$.
        \ENDFOR
    \ENDFOR
\ENDFOR
\FOR{each pair of ICPs $(p_i, p_j)$}
    \STATE Find the common KPIs between $p_i$ and $p_j$ using $param\_to\_kpis$.
    \IF{common KPIs exist}
        \STATE Add an edge between $p_i$ and $p_j$ in $G$.
        \STATE Label the edge with the common KPIs.
    \ENDIF
\ENDFOR
\FOR{each KPI $k$}
    \STATE Find all ICPs associated with $k$.
    \STATE Group these ICPs in $P_k^G$.
\ENDFOR
\RETURN $P_k^G$.
\end{algorithmic}
\label{algo:param_kpi_grouping}
\end{algorithm}

When the conflict is detected, the \ac{CDC} signals the \ac{CMC} about the specific types of conflict occurring over the specific conflicting \ac{ICP}. For the analysis of this paper, we adopt the priority conflict mitigation method proposed by Adamczyk et al. \cite{adamczyk2023conflict} and \ac{QACM} method proposed in \cite{wadud2024qacm} to be deployed in the \ac{CMC}. The priority method simply priorities one conflicting \ac{xApp} over the other based on their significance set by the \ac{MNO}. On the other hand, the \ac{QACM} considers specific \ac{QoS} requirements of the conflicting \acp{KPI} and suggests an optimal value of the conflicting parameter at which the \acp{KPI} are either above their \ac{QoS} mandates or as closer as possible to it. The following section discusses the simulation setup and results based on these two mitigation methods. 


\begin{table}[h]
    \centering
    \caption{Mobility and Service Type Distribution}
    \label{tab:mobility_distribution}
    \renewcommand{\arraystretch}{1}
    \resizebox{\linewidth}{!}{ 
    \small
    \begin{tabular}{|l|l|}
        \hline
        \textbf{Feature} & \textbf{Description} \\ 
        \hline
        Mobility Models & Walking (0-1 m/s), Cycling (2-5 m/s), Driving (6-15 m/s) \\ 
        \hline
        UE Mobility Distribution & (35\%) Walking, (30\%) Cycling, (35\%) Driving \\ 
        \hline
        Service Types & eMBB (40\%), URLLC (30\%), mMTC (30\%) \\ 
        \hline
        Movement Logic & Random 2D movement with boundary reflection \\ 
        \hline
    \end{tabular}}
\end{table}

\section{Experiment and Results}
\label{sec:simRes}
For the experimental analysis, we used MATLAB software for simulation, along with its 5G Toolbox and O-RAN 7.2 split \cite{arafat2024transformer, wadud2024qacm} and conducted all simulations on a MacBook Pro with an M1 chip and 8GB of RAM. The simulation setup included four \acp{gNB} and hundred \acp{UE}. We focused on a direct conflict between the \ac{MRO} \ac{xApp} and \ac{ES} \ac{xApp} over \ac{TXP} values. The mobility and service type distribution is shown in Table~\ref{tab:mobility_distribution}. The simulation parameters were: a \ac{RSRP} threshold of -110 dBm for handover, a frequency of 2.4 GHz, and a simulation time of 10 minutes with a time step of 100 ms. Input control parameters included a default \ac{TXP} value of 30 dBm, \ac{CIO} of 2 dB, \ac{HYS} of 0.5 dB, \ac{TTT} of 0.1 ms, \ac{RET} of 1.5 degrees, and an adjustment interval of 1000 ms. The \acp{UE} moved back and forth between the two \acp{gNB} with randomly assigned velocities between 0 and 5 m/s. The overall system goal is to minimize link failures and ping-pong handovers for \ac{MRO} while maximizing energy efficiency for \ac{ES}. A direct conflict arose when \ac{ES} set the \ac{TXP} to 3 dBm for energy savings, followed by \ac{MRO} setting it to 50 dBm in the next request. We consider 5 conflict mitigation Methods, including a no-mitigation approach (NC) where the \ac{TXP} was set to the recently requested value; a set back to default (SBD) method where the \ac{TXP} was reset to its default value; a priority-based method where either \ac{ES} (P-ES) or \ac{MRO} (P-MRO) was prioritized; and finaly the \ac{QACM} method was used to balance the \ac{TXP} based on \ac{QoS} thresholds. We repeated the simulation 500 times and the results were presented using box plots for each \ac{KPI}.

\begin{figure*}
  \centering
  \hfill
  \begin{subfigure}[b]{0.32\textwidth}
 \centering
 \includegraphics[width=\textwidth]{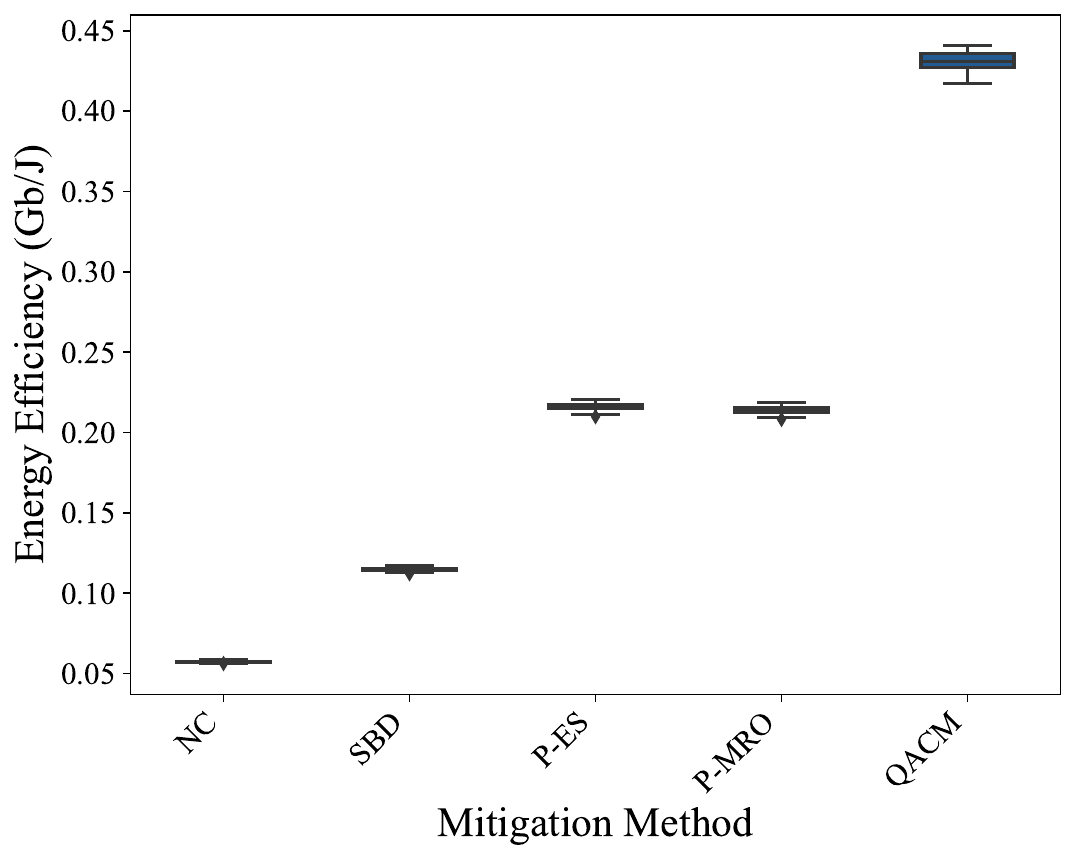}
 \caption{Energy efficiency.}
 \label{fig:energyEff}
  \end{subfigure}
  \hfill
  \begin{subfigure}[b]{0.32\textwidth}
 \centering
 \includegraphics[width=\textwidth]{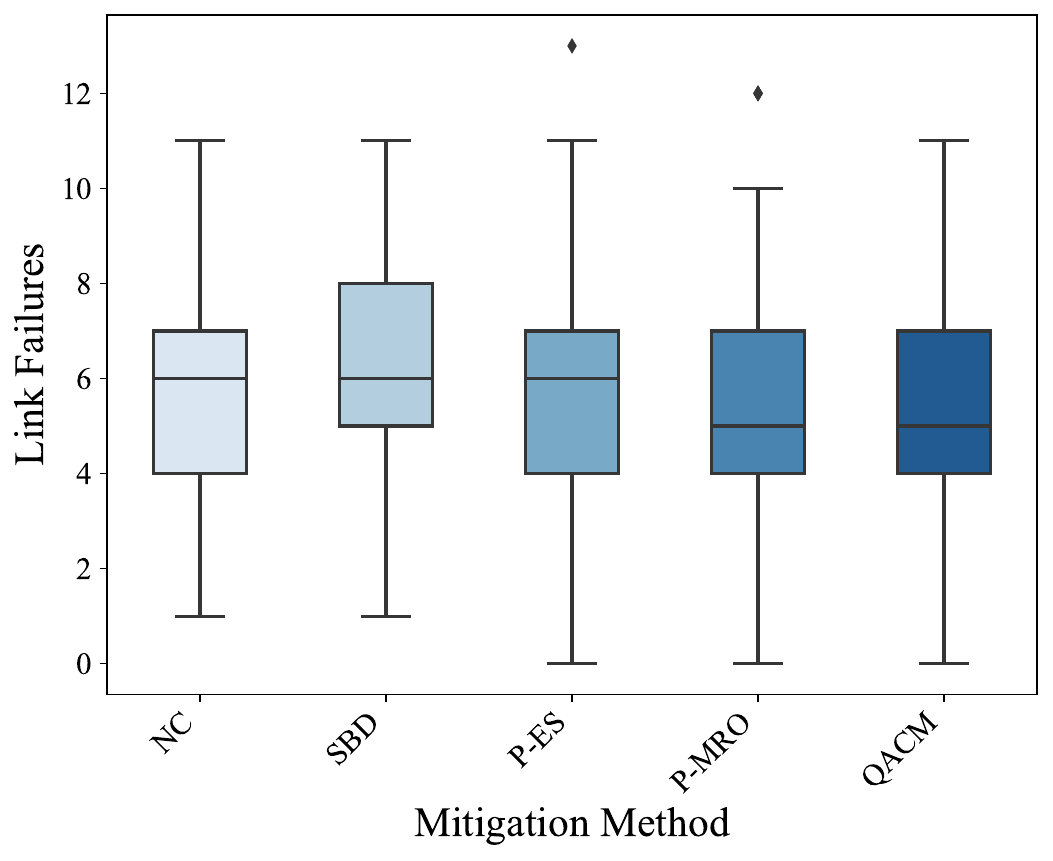}
 \caption{Number of link failures.}
\label{fig:linkFail}
\end{subfigure}
\hfill
\begin{subfigure}[b]{0.33\textwidth}
 \centering
 \includegraphics[width=\textwidth]{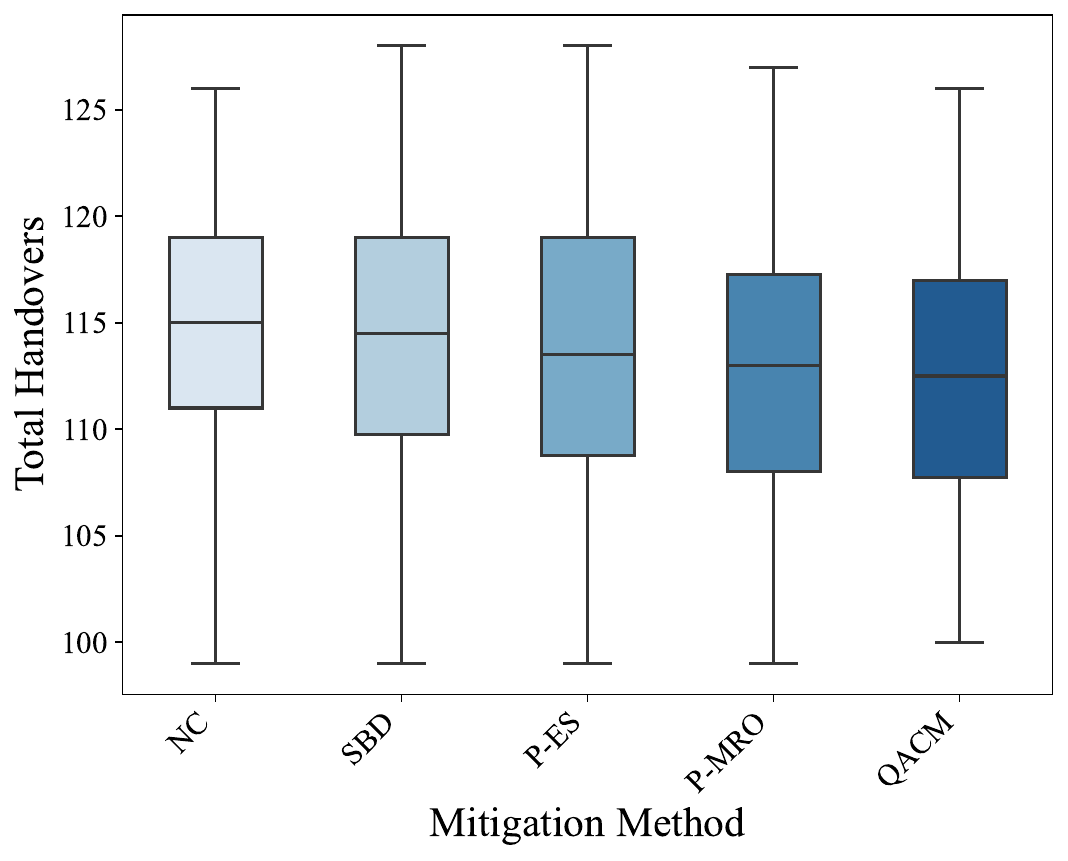}
 \caption{Number of total handovers.}
	\label{fig:totalHandover}
\end{subfigure}
  \caption{Simulation results of \acp{KPI} adopting different conflict mitigation methods.}
 \label{fig:res}
  \vspace{-0.3in}
\end{figure*}


The Fig.~\ref{fig:energyEff} compares the energy efficiency across different conflict mitigation methods, including NC (No Conflict Mitigation), SBD (Set Back to Default), P-ES (prioritized ES), P-MRO (prioritized MRO), and QACM. The QACM method shows the highest energy efficiency (in bits/Joule), suggesting it is the most effective approach for maximizing energy savings, followed by P-ES. The NC, SBD and P-MRO methods exhibit lower energy efficiency indicating that prioritizing energy-saving strategies improves this KPI. The \ac{QACM} optimally sets the transmission power to minimize link failures and unnecessary ping-pong handovers, as reflected in Fig.~\ref{fig:energyEff}. \ac{QACM} demonstrates 90-100\% greater energy efficiency compared to P-\ac{ES} or P-\ac{MRO}. This is because none of the prioritized methods effectively balance the transmitted bits-to-energy ratio while achieving their own objectives. SBD and NC exhibit the second-lowest and lowest performance, respectively, as the default values do not optimize any \ac{xApp}'s objectives. On the other hand, maintaining the value demanded by \ac{MRO} negatively impacts energy efficiency, since this \ac{KPI} is not associated with \ac{MRO} but rather with \ac{ES}. Therefore, \ac{MRO}'s instruction doesn't assure positive impact on energy saving. 


The Fig.~\ref{fig:linkFail} illustrates the impact of different conflict mitigation methods on reducing link failures. The SDB displays higher numbers of link failures followed by the NC and P-ES, with having a similar median value, implying poorer performance in maintaining stable connections. On the other hand, P-MRO and QACM show significantly fewer link failures with QACM being the most effective one. This indicates that prioritizing mobility robustness and a balanced conflict resolution approach reduces the occurrence of link failures. Overall, the \ac{QACM} reduces approximately 17\% of connection losses compared to the NC method. 




Finally, the Fig.~\ref{fig:totalHandover} demonstrates the total number of handovers across the various methods. NC, SBD and P-ES have the highest frequency of handovers, which may negatively affect network stability. In contrast, P-MRO and QACM exhibit significantly fewer handovers with QACM providing the best mitigation. This shows that a mobility-optimized and balanced conflict resolution strategy can effectively reduce these undesirable handovers. Overall, the \ac{QACM} saves approximately 2.2\% of unnecessary handovers compared to the NC method.

\section{Limitations and Future Works}
\label{sec:limitF}
This study relies on a MATLAB-based simulation with a scenario of one direct conflict between MRO and ES, and a small-scale network of four gNBs and one hundred UEs. This setup may not fully capture real-world complexities such as OTA (Over The Air) interference, multi-cell topologies, or stringent latency demands. In this work, we primarily focused on energy efficiency and did not explore multi-objective or ML-based strategies. This certainly leaves questions about scalability, signaling overhead, and adaptive response of the proposed solutions in rapidly changing environments. Our future work will include deploying on more realistic simulator like ns3-oran \cite{ns3oran_github} developed by NIST, where we have been actively developing and deploying xApps, investigating more diverse conflicts, incorporating multiple objectives, and exploring ML-driven methods for proactive conflict mitigation.

\section{Conclusion}
\label{sec:conclusion}
In this paper, we provided an in-depth examination of \ac{xApp}-level conflicts in Open \ac{RAN} focusing on conflict detection and mitigation strategies. Through theoretical modeling and the use of conflict graphs, we identified different types of conflicts and highlighted their impacts on network \acp{KPI}. Our simulation results demonstrated the effectiveness of various mitigation methods, particularly the \ac{QACM} approach, in managing conflicts while respecting \ac{QoS} thresholds. The results showed that \ac{QACM} outperforms other mitigation strategies, including no conflict mitigation, reset to default and priority-based approaches in terms of energy efficiency and minimizing link failures and total number of handovers (Including ping-pong handovers). Our future work will focus on refining conflict detection mechanisms experimenting with the detection mechanism in large-scale, exploring machine learning-based mitigation techniques, and evaluating real-world O-RAN deployments to further enhance the robustness of conflict management in dynamic network environments.


%



\section*{Acknowledgment}
This research was partially funded by the European Union's Horizon Europe research and innovation program through the Marie Sklodowska-Curie SE grant, under agreement number RE-ROUTE No 101086343.

\ifCLASSOPTIONcaptionsoff
  \newpage
\fi




\begin{thebibliography}{10}

\bibitem{adamczyk2023conflict}
C.~Adamczyk and A.~Kliks, ``Conflict mitigation framework and conflict detection in o-ran near-rt ric,'' {\em IEEE Communications Magazine}, vol.~61, no.~12, pp.~199--205, 2023.

\bibitem{wadud2023conflict}
A.~Wadud, F.~Golpayegani, and N.~Afraz, ``Conflict management in the near-rt-ric of open ran: A game theoretic approach,'' in {\em 2023 IEEE International Conferences on Internet of Things (iThings) and IEEE Green Computing \& Communications (GreenCom) and IEEE Cyber, Physical \& Social Computing (CPSCom) and IEEE Smart Data (SmartData) and IEEE Congress on Cybermatics (Cybermatics)}, pp.~479--486, IEEE, 2023.

\bibitem{wadud2024qacm}
A.~Wadud, F.~Golpayegani, and N.~Afraz, ``Qacm: Qos-aware xapp conflict mitigation in open ran,'' {\em IEEE Transactions on Green Communications and Networking}, vol.~8, no.~3, pp.~978--993, 2024.

\bibitem{erdol2024xapp}
H.~Erdol, X.~Wang, R.~Piechocki, G.~Oikonomou, and A.~Parekh, ``xapp distillation: Ai-based conflict mitigation in b5g o-ran,'' {\em arXiv preprint arXiv:2407.03068}, 2024.

\bibitem{del2024pacifista}
P.~B. del Prever, S.~D'Oro, L.~Bonati, M.~Polese, M.~Tsampazi, H.~Lehmann, and T.~Melodia, ``Pacifista: Conflict evaluation and management in open ran,'' {\em arXiv preprint arXiv:2405.04395}, 2024.

\bibitem{ric_oran_alliance}
O.-R. Alliance, ``O-ran working group 3 (near-real-time ran intelligent controller and e2 interface workgroup), near-rt ric architecture,'' {\em O-RAN.WG3.RICARCH-R003-v04.00}, Last Accessed [September 2024].

\bibitem{polese2022understanding}
M.~Polese, L.~Bonati, S.~D’Oro, S.~Basagni, and T.~Melodia, ``Understanding o-ran: Architecture, interfaces, algorithms, security, and research challenges,'' {\em IEEE Communications Surveys \& Tutorials}, vol.~25, no.~2, pp.~1376--1411, 2023.

\bibitem{adamczyk2023detection}
C.~Adamczyk and A.~Kliks, ``Detection and mitigation of indirect conflicts between xapps in open radio access networks,'' in {\em IEEE INFOCOM 2023-IEEE Conference on Computer Communications Workshops (INFOCOM WKSHPS)}, pp.~1--2, IEEE, 2023.

\bibitem{corici2024towards}
M.~Corici, R.~Modroiu, F.~Eichhorn, E.~Troudt, and T.~Magedanz, ``Towards efficient conflict mitigation in the converged 6g open ran control plane,'' {\em Annals of Telecommunications}, pp.~1--11, 2024.

\bibitem{adamczyk2023challenges}
C.~Adamczyk, ``Challenges for conflict mitigation in o-ran’s ran intelligent controllers,'' in {\em 2023 International Conference on Software, Telecommunications and Computer Networks (SoftCOM)}, pp.~1--6, IEEE, 2023.

\bibitem{zhang2022team}
H.~Zhang, H.~Zhou, and M.~Erol-Kantarci, ``Team learning-based resource allocation for open radio access network (o-ran),'' in {\em ICC 2022-IEEE International Conference on Communications}, pp.~4938--4943, IEEE, 2022.

\bibitem{arafat2024transformer}
M.~Arafat~Habib, P.~E. Iturria-Rivera, Y.~Ozcan, M.~Elsayed, M.~Bavand, R.~Gaigalas, and M.~Erol-Kantarci, ``Transformer-based wireless traffic prediction and network optimization in o-ran,'' {\em arXiv e-prints}, pp.~arXiv--2403, 2024.

\bibitem{zolghadr2024learning}
A.~Zolghadr, J.~F. Santos, L.~A. DaSilva, and J.~Kibi{\l}da, ``Learning and reconstructing conflicts in O-RAN: A graph neural network approach,'' {\em arXiv e-prints}, pp.~arXiv--2412.14119, 2024.

\bibitem{ns3oran_github}
U.S. National Institute of Standards and Technology (NIST), ``ns3-oran,'' {\em GitHub Repository}, 2024. [Online]. Available: \url{https://github.com/usnistgov/ns3-oran} [Accessed: September 15, 2024].

\end{thebibliography}
%
\bibliographystyle{ieeetr}
\end{document}